\documentclass{article}
\newcommand{\be}{\begin{equation}}
\newcommand{\ee}{\end{equation}}
\newcommand{\bt}{\beta}
\newcommand{\fr}{\frac}
\begin{document}
\baselineskip=16pt
\begin{center}
{\large\bf Cosmic Acceleration With A Positive Cosmological Constant}\\
\vspace{1.2cm} {\bf \large Arbab I. Arbab}\footnote{On leave from
Comboni College for Computer Science, P.O. Box 114, Khartoum,
Sudan\\ } \footnote{E-mail: arbab@ictp.trieste.it}
\\ \vspace{.5cm}
Department of Physics, Teacher's College, Riyadh 11491, P.O.Box
4341, Kingdom of Saudi Arabia
\end{center}
\vspace{.6in}
We have considered a cosmological model with a
phenomenological model for the cosmological constant of the form
$\Lambda=\bt\fr{\ddot R}{R}$, $\ \bt$ is a constant. For age
parameter consistent with observational data the Universe must be
accelerating in the presence of a positive cosmological constant.
The minimum age of the Universe is $H_0^{-1}$, where $H_0$ is the
present Hubble constant.  The cosmological constant is found to
decrease as $t^{-2}$. Allowing the gravitational constant to
change with time leads to an ever increasing gravitational
constant at the present epoch. In the presence of a viscous fluid
this decay law for $\Lambda$ is equivalent to the one with
$\Lambda=3\alpha H^2$ ($\alpha=\rm const.$) provided
$\alpha=\fr{\bt}{3(\bt-2)}$. The inflationary solution obtained
from this model is that of the de-Sitter type.
\\
\vspace{0.3cm}
\vspace{1cm}\\

{\bf 1. INTRODUCTION}\\
\baselineskip=20pt One of the puzzling problems in standard
cosmology is the cosmological constant problem. Observational data
indicate that $\Lambda\sim 10^{-55}\rm cm^{-2}$ while particle
physics prediction for $\Lambda$ is greater than this value by a
factor of order  $10^{120}$. This discrepancy is known as the
cosmological constant problem. A point of view which allows
$\Lambda$ to vary in time is adopted by several workers. The point
is that during the evolution of the universe the energy density of
the vacuum decays into particles thus leading to the decrease of
the cosmological constant. As a result one has the creation of
particles although the typical rate of the creation is very small.
The entropy problem which exists in the Standard Model can be
solved by the above mechanism. One of the motivations for
introducing $\Lambda$ term is to reconcile the age parameter and
density parameter of the universe with current observational data.
Recent observations of Type 1a supernovae which indicate an
accelerating universe draw once more the attention to the possible
existence, at the present epoch, of a small positive cosmological
constant ($\Lambda$). One possible cause of the present
acceleration could be the ever increasing gravitational (constant)
forces. As a consequence, a flat universe has to speed up so that
gravitational attraction should not win over expansion. Or
alternatively, the newly created particles give up their kinetic
energy to push the expansion further away.

The purpose of this work is to study the phenomenological decay
law for $\Lambda$ that is  proportional to the deceleration
parameter. In an attempt to modify the general theory of
relativity, Al-Rawaf and Taha [17] related the cosmological
constant to the Ricci scalar, $\cal{R}$. This is written as a
built-in cosmological constant, i.e., $\Lambda\propto\cal{R}$. A
comparison with our ansatz above for $\Lambda$ yields a similar
behaviour for a flat universe. And since Ricci scalar contains a
term of the form $\fr{\ddot R}{R}$, one adopts this variation for
$\Lambda$. We parameterized this as $\Lambda=\bt\fr{\ddot R}{R}$,
where $\bt$ is a constant. The cosmological consequences of this
decay law are very attractive. This law finds little attention
among cosmologists. This law provides a reasonable solution to
the cosmological puzzles presently known. We have found that a
resolution to these problems is possible with a positive
cosmological constant ($\Lambda >0$). This requires the
deceleration parameter to be negative ($q< 0$). Usually people
invoke some kind of a scalar field that has an equation of state
of the form $p< 0,\ p$ the pressure of the scalar field. A more
recent review for the case of a
positive cosmological constant is found in [6].\\
A variable gravitational constant $G$ can also be incorporated into a simple
framework in which $\Lambda$ varies as well, while retaining the usual 
energy
conservation law [1,10,11]. The above decay law leads to a power-law 
variation for $G$.
Inflationary solutions are also possible with this mechanism thus solving 
the
Standard Model problems. We have recently shown that a certain variation of 
$G$
may be consistent with palaeontological as well as geophysical data [15].
\\

{\bf 2. THE MODEL}\\
For the Friedmann--Robertson-Walker metric the Einstein's field
equations with the variable cosmological constant and  a source
term given by a stress-energy tensor of a perfect fluid read \be
3\fr{\dot R^2}{R^2}+\fr{3k}{R^2}=8\pi G\rho+\Lambda\ , \ee \be
2\fr{\ddot R}{R}+\fr{\dot R^2}{R^2}+\fr{k}{R^2}=-8\pi Gp+\Lambda
\ee where $\rho$ is the fluid energy density and $p$ its pressure.
The equation of the state is taken in the  form \be
p=(\gamma-1)\rho \ee where $\gamma$ is a constant. From eqs.(1)
and (2) one finds \be \fr{d(\rho
R^3)}{dt}+p\fr{dR^3}{dt}=-\fr{R^3}{8\pi G}\fr{d\Lambda}{dt} \ee We
propose a phenomenological decay law for $\Lambda$ of the form
[4,5] \be \Lambda=\bt\fr{\ddot R}{R} \ee where $\bt$ is a
constant. Overdin and Cooperstock have pointed out that there is
no fundamental difference between the first and second derivatives
of the scale factor that would preclude the latter from acting as
an independent variable if the former is acceptable [4]. Moreover,
from eqs.(1) and (4), one can write \be \ddot R=\fr{8\pi
G}{3}\left(1-\fr{3\gamma}{2}\right)\rho R+\fr{\Lambda}{3}R\ . \ee
Thus one example of $\Lambda$ in the above form is the case when
the universe is filled with a fluid characterized by
$\gamma=\frac{2}{3}$ and $\beta=3$. For other values of $\gamma
(=1)$, $\beta$ is not constrained by the Einstein equations, and
the general relation \begin{equation}
\Lambda=\left(\frac{\beta}{\beta-3}\right)4\pi G \rho
\end{equation}
shows the ratio of $\Lambda$ to $\rho$ is constant in this
phenomenological model. Now eq.(1) together with eq.(7) yield
\begin{equation}
\Lambda=\left(\frac{\beta}{\beta-2}\right)H^2.
\end{equation}
Thus as remarked by Overdin and Cooperstock the model with
$\Lambda\propto H^2$ and the above form (eq.(5)) are basically
equivalent. We see that in the radiation and in the matter
dominated eras the vacuum contributes significantly to the total
energy density of the universe in both eras. Thus unless the
vacuum always couples (somehow) to gravity such a behavior can not
be guaranteed (and understood) at both epochs. Such a mechanism is
exhibited in eq.(20). Hence the vacuum domination of the present
universe is not accidental but a feature that is present at all
times. One would expect that there must have been a conspiracy
between the two components in such a way the usual energy
conservation holds. Therefore, one may argue that in cosmology the
energy conservation principle is not a {\it priori} principle
[17]. We observe from eq.(7) that $\frac{\Lambda_0}{\Lambda_{\rm
Pl}}\approx \frac{\rho_0}{\rho_{\rm
Pl}}\approx\left(\frac{10^{-29}}{10^{93}}\right)=10^{-122}$, where
"0" refers to the present  and "Pl" refers to Planck era of the
quantity, respectively. Thus such a phenomenological model for
$\Lambda$ could provide a {\it natural} answer (interpretation) to
the puzzling question why the cosmological constant is so small
tody, rather than just attributing it to the oldness of our
present universe. \\
For the matter-dominated universe $\gamma=1$
and therefore eqs.(2),(3), and (5) yield (for $k=0$) \be
(\bt-2)\ddot RR=\dot R^2 \ , \ee which can be integrated to give
\be R(t)=\left(\fr{A(\bt-3)}{(\bt-2)}t\right)^{(\bt-2)/(\bt-3)} \
,\ \ \bt\ne 3\ , \ \  \bt\ne 2\ , \ee where $A=\rm consatnt$. It
follows from eq.(5) that \be
\Lambda(t)=\fr{\bt(\bt-2)}{(\bt-3)^2}\fr{1}{t^2}\ \ \ ,\ \ \bt\ne
3\ . \ee Using eqs.(1), (5) and (10) the energy density can be
written as, \be \rho(t)=\fr{(\bt-2)}{(\bt-3)}\fr{1}{4\pi G\ t^2} \
\ , \ \bt\ne 3\  \ee and the vacuum energy density ($\rho_v$) is
given by \be \rho_v(t)=\fr{\Lambda}{8\pi
G}=\fr{\bt(\bt-2)}{(\bt-3)^2}\fr{1}{8\pi G t^2}\ \  , \ \ \bt\ne
3\ . \ee The deceleration parameter ($q$) is defined as \be
q=-\fr{\ddot RR}{\dot R^2}=\fr{1}{2-\bt}\ , \ \bt\ne 2\ . \ee We
see from eqs.(11) and (14) that for a positive $\Lambda$ the
deceleration parameter is negative (for $\bt > 2$). For $\bt<2$
the cosmological constant is negative, $\Lambda<0$. It has been
recently found that the universe is probably accelerating at the
present epoch. There are several justifications for this
acceleration. Some authors attribute this acceleration to the
presence of some scalar field (quintessence field) with a negative
pressure filling the whole universe. And this field has a
considerable contribution to the total energy density of the
present universe. The density parameter of the universe
($\Omega_m$) is given by \be
\Omega_m=\fr{\rho}{\rho_c}=\fr{2}{3}\fr{(\bt-3)}{(\bt-2)}\ \ , \
\bt\ne 2 \ee
where $\rho_c=\fr{3H^2}{8\pi G}$ is the critical energy density of the
universe.
We notice that the Standard Model formula $\Omega_m=2q$ is now replaced by
$\Omega_m=\fr{2}{3}q+\fr{2}{3}$. However, both models give $q=\frac{1}{2}$ 
for
a critical density. This relation has been found by several authors [1,3].
\\
The density parameter due to vacuum contribution is defined as
$\Omega^\Lambda=\fr{\Lambda}{3H^2}$. Employing eqs.(8) this gives
\be \Omega^\Lambda=\fr{\bt}{3(\bt-2)}\ , \ \ \bt\ne 2\ . \ee We
shall define $\Omega_{\rm total}$ as \be \Omega_{\rm
total}=\Omega_m+\Omega^\Lambda \ee Hence eqs.(1), (7), (15) and
(16) give $\Omega_{\rm total}=1$. This setting is favored by the
inflationary scenario. The present value of the age of the
universe, deceleration parameter and the cosmological constant are
obtained from eqs.(10), (11) and (15) \be
t_0=\fr{(\bt-2)}{(\bt-3)}H_0^{-1} ,\ \
\Omega_{m0}=\fr{2}{3}\fr{(\bt-3)}{(\bt-2)} \ , \ \
\Lambda_0=\fr{\bt}{(\bt-2)}H_0^2\ \ , \ \ \bt\ne 2,\ \ \bt\ne 3\ .
\ee (the subscript `0' denotes the present value of the quantity
and $H$ is Hubble constant). Inasmuch as $q<0$, $\Omega_m<1$ hence
the low density of the universe is no longer a problem. Moreover,
in order to solve the age problem we require $\beta>3$. Thus the
constrain $\beta>3$ may resolve both problems. Observational
evidence, however, does not rule out the negative deceleration
parameter and the stringent limits on the present value
of $q_0$ are $-1.25\le q_0\le 2$ [8].\\
The case $\bt=2$ represents an empty static universe with
$\Lambda=0$. For $\beta=0$,  the usual expressions for FRW models
are recovered. When $\beta=0$, $\Omega_{m0}=1$, which is {\it
in}consistent with many observational tests on scales much too
small to be affected by the cosmological constant, e.g. dynamical
tests for scales up to a few tens of Mpc. For $\bt=4$ one obtains
$t_0=2H_0^{-1}, \Omega_{m0}=\fr{1}{3},
\Omega_0^\Lambda=\fr{2}{3}$; $\bt=12$, $t_0=\fr{10}{9}H_0^{-1},
\Omega_{m0}=\fr{3}{5}, \Omega_0^\Lambda=\fr{2}{5}$. It is
interesting to note that when $\bt\rightarrow \infty $, all
parameters are finite, namely, $R_0\rightarrow t_0$,
$t_0\rightarrow H_0^{-1}, \Lambda\rightarrow H_0^2$,
$\Omega_{m0}\rightarrow \fr{2}{3}, \Omega_0^\Lambda=\fr{1}{3}$.
Thus the values of $\beta$ which are consistent with $$
\Omega_{m0}=0.3\pm 0.1 $$ are $$ \beta=4.0\pm 0.5 $$ but the ages
are very high. For instance, with a high, but still consistent
with observational constraints, value of the Hubble constant,
$H_0=80\ \rm km/s/Mpc$, and a rather generous, $$ t_0=15\pm 2 \ \
\rm Gyr$$ then values of $\beta$ consistent with this are:
$$ 5.5 < \beta \le 19\  .$$
This suggests a best fit value somewhere around $\beta=5$, which
would give the somewhat high matter density of $\Omega_{m0}=0.44$
and rather high age of $t_0=18.3\ \rm Gyr$. We have recently [15]
investigated the implications of a variable $G$ on the Earth-Sun
system, contrary to what have been believed, we have found that
the palaeontological data are consistent with a variable $G$
provided that the age of the universe is $t_0\sim 11\times10^{9}$
years and that $G\propto t^{1.3}$. However, a recent value for
the age of the universe from gravitational lensing suggests
$t_0\sim 11\times10^{9}$ years. Recent estimates from
observations of galaxy clustering and their dynamics indicate
that the mean mass density is about one-third of the critical
value [9]. Thus if $\bt\ne0$ then the present age of the universe
can not be less than $H_0^{-1}$. This constrain represents our
strongest prediction for the age of the Universe.
\\
\\
{\bf 3. A MODEL WITH VARIABLE $G$}\\
We now consider a model in which both $G$ and $\Lambda$ vary with
time. Imposing the usual energy conservation law one obtains
[1,10,11] \be \dot\rho+3\gamma H\rho=0\ , \ee and \be
\dot\Lambda+8\pi\dot G\rho=0\ . \ee Using eqs.(10) and (11),
eqs.(19) and (20) yield \be \rho(t)=Dt^{-3(\bt-2)/(\bt-3)}\ , \ \
D=\rm const. \ee and \be G(t)=\left (\fr{(\bt-2)}{4\pi
A(\bt-3)}\right)t^{\bt/(\bt-3)}\ \, \bt\ne 3\ . \ee For $\bt=0$,
$G$=const. and $\rho=Dt^{-2}$ and
$R=\left(\fr{3}{2}At\right)^{2/3}$, which is the usual FRW result.
Clearly for $\bt>3$ the gravitational constant increases with time
while for $\bt<3$ it decreases with time. Once again the
constraint $\bt>3$ considered before implies that the
gravitational constant increases with time. An increasing
gravitational constant is
considered by several workers [1,10,11,14].\\
In a recent work we have shown that the variation of the gravitational
constant is consistent with palaeontological data [15].
The gravitational constant might have had a very different value
from the present one. This depends strongly on the value of $\bt$ assumed
at a given epoch. The development of the large-scale anisotropy is given
by the ratio of the shear $\sigma$ to the volume expansion
($\theta=3\fr{\dot R}{R}$) which evolves as [16]
\be
\fr{\sigma}{\theta}\propto t^{(3-2\bt)/(\bt-3)}\ ,
\ee
and since $\bt>3$ this anisotropy decreases as the universe expands
and this explains the present observed isotropy of the Universe.\\
For example, if $\bt=\fr{3}{2}$ then $G\propto t^{-1}, R\propto
t^{1/3}, \rho\propto t^{-1}$. This behavior of $G$ was considered
by Dirac in his Large Number Hypothesis (LNH) model [12]. In an
earlier work we have shown that some non-viscous models are
equivalent to bulk viscous ones. This behavior is also manifested
in our present model provided one takes
$\bt=\fr{3(2n-1)}{(3n-2)}$, where the bulk viscosity ($\eta$) is
defined as $\eta=\rm const.\rho^n$, where $0\le n\le 1$ [1]. The
two models, though different in the form of $\Lambda$, are
equivalent if one puts $\alpha=\fr{\bt}{3(\bt-2)}$. Thus the decay
law $\Lambda=\bt\fr{\ddot R}{R}$ and $\Lambda=3\alpha H^2$, where
$\alpha=\rm const.$, are identical in the presence of a cosmic
fluid.
\\
\\
{\bf 4. AN INFLATIONARY SOLUTION }  \\
This solution is obtained from eqs.(1), (2), (3) and (5) with
$\bt=3$. One then gets $R\ddot R=\dot R^2$, which integrated to
give \be R=\rm const.\exp(Ct) \ee where $H=C=\rm const.$. Applying
eq.(24) to eqs.(5) and (20), employing eq.(1), we get \be
\Lambda=3H^2\ \ , \ \ \rho=0 \ \ , \ \ G=\rm const.\ . \ee This is
the familiar de-Sitter inflationary solution (in the matter
dominated epoch). A similar inflationary solution is obtained with
$\bt=3$ in the radiation dominated epoch. Inflationary models
employ a scalar field (inflaton) to arrive at this solution. These
solutions help resolve several cosmological problems associated
with the standard model (flatness, horizon, monopole, etc.) The
inflationary solution resolves some of the outstanding issues of
standard cosmology.
\\
\\
{\bf 5. CONCLUSION}\\
In this paper we have considered the cosmological implications of
a decay law for $\Lambda$ that is proportional to $\frac{\ddot
R}{R}$. The model is found to be very interesting and apparently a
lot of problems can be solved. To solve the age parameter and the
density parameter one requires the cosmological constant to be
positive or equivalently the deceleration parameter to be
negative. This implies an accelerating universe. However, the
strongest support for an accelerating universe comes from
intermediate redshift results for Type 1a supernovae. The model
predicts that the minimum age of the Universe is $H_0^{-1}$. The
behavior that $\Lambda\propto t^{-2}$ is found by several authors.
The gravitational constant is found to increase with time at the
present epoch. Our model predicts an inflationary phase in the
matter dominated epoch as well as in the radiation dominated
epoch. The cosmological tests for this model can be obtained from
those ones already investigated by us [7]. The choice among these
models awaits the emergence of the new data.
\\
\\
\\
\\
{\bf ACKNOWLEDGEMENTS} \\
My ideas on this subject have benefited from discussions with a
number of friends and colleagues. I am grateful to all of them. I
wish to thank Omdurman Ahlia University for financial support of
this work. I would like to thank the anonymous referees for their
useful comments and corrections.
\\
\\
{\bf REFERENCES}\\
1- Arbab, A. I., 1997. {\it Gen. Rel. Gravit.}{\bf 29}, 61\\
2- Beesham, A., 1993. {\it Phys. Rev.} {\bf D48}, 3539\\
3- Matyjasek, J., 1995. {\it Phys. Rev.}{\bf D51}, 4154\\
4- Overdin, J. M., and Cooperstock, F.I., 1998. {\it Phys. Rev.}{\bf
D58}, 043506\\
5- Al-Rawaf, A. S., 1998. {\it Mod. Phys. Lett.}{\bf A 13}, 429\\
6- Sahni, V., and Starobinsky, A., 1999. {\it Los Alamos preprint}
astro-ph/9904398 \\
7- Arbab, A. I., 1998. {\it Astrophys. Space Science}{\bf 259}, 371\\
8- Klapdor, H. W., and Grotz, K., 1986. {\it Astrophys. J.}{\bf 301}, l39\\
9- Peebles, P. J. E., 1986. {\it Nature }{\bf 321}, 27\\
10- Abdel-Rahman, A. -M. M., 1990. {\it Gen. Rel. Gravit.}{\bf 22}, 655\\
11- Beesham, A., 1986. {\it Nouvo Ciment.}{\bf B96}, 17\\
12- Dirac, P. A. M., 1937. {\it Nature }{\bf 139}, 323\\
13- Kalligas, D., Wesson, P., and Everitt, C. W., 1992. {\it Gen. Rel.
Gravit.}{\bf 24}, 351\\
14- Abdussattar, A., and Vishwakarma, R. G., 1997. {\it Class.
Quantum
Gravit.}{\bf 14}, 945\\
15- Arbab, A. I., 1998. {\it Los Alamos preprint} physics/9811024\\
16- Barrow, J. D., 1978. {\it Mon. Not. astr. Soc. 184}, 677\\
17- Al-Rawaf, A.S., and Taha, M.O., 1996. {\it Gen. Rel.
Gravit.}{\bf 28}, 935,
\\
\end{document}